\documentclass{iopart}

\usepackage{xspace}
\usepackage{theorem}

\newcommand{\tO}{\ensuremath{t_{\rm o}}\@\xspace}
\newcommand{\tE}{\ensuremath{t_{\rm e}}\@\xspace}

\newcommand{\tOm}{\ensuremath{t_{\rm o}}}
\newcommand{\tEm}{\ensuremath{t_{\rm e}}}

\newcommand{\tauOt}{\ensuremath{\tau_{\rm o}}\@\xspace}
\newcommand{\tauEt}{\ensuremath{\tau_{\rm e}}\@\xspace}
\newcommand{\tauO}{\ensuremath{\tau_{\rm o}}}
\newcommand{\tauE}{\ensuremath{\tau_{\rm e}}}

\newcommand{\TEt}{\ensuremath{T_{\rm e}}\@\xspace}

\newcommand{\TE}{\ensuremath{T_{\rm e}}}

\newcommand{\Io}{{\bf I$_1$}\@\xspace}
\newcommand{\AIo}{{\bf AI$_1$}\@\xspace}
\newcommand{\It}{{\bf I$_2$}\@\xspace}
\newcommand{\AIt}{{\bf AI$_2$}\@\xspace}

\theoremstyle{break} \newtheorem{theo}{Theorem}

\begin{document}

%%%%%%%%%%%%%%%%%%%%%%%%%%%%%%%%%
%%% Classical and quantum gravity
%%%%%%%%%%%%%%%%%%%%%%%%%%%%%%%%%
\jl{6}

%%%%%%%%%%%%%%%%%%%%%%%%%%
%%% If we make it a letter
%%%%%%%%%%%%%%%%%%%%%%%%%%
\letter{Anisotropic universes with isotropic cosmic microwave
  background radiation}

%%%%%%%%%%%%%%%%%%%%%%%%%
%%% If we make it a paper
%%%%%%%%%%%%%%%%%%%%%%%%%
%%  \title{Anisotropic universes with isotropic cosmic microwave
%%  background radiation}

%%%%%%%%%%%%%%%%%%%%%%%%%
%%% Authors and addresses
%%%%%%%%%%%%%%%%%%%%%%%%%
\author{W C Lim, U S Nilsson and J Wainwright}
\address{Department of Applied Mathematics\\
  University of Waterloo\\
  Waterloo, Ontario\\
  Canada N2L 3G1}

%%%%%%%%%%%%
%%% Abstract
%%%%%%%%%%%%
\begin{abstract}
  We show the existence of spatially homogeneous but anisotropic
  cosmological models whose cosmic microwave background temperature is
  exactly isotropic at one instant of time but whose rate of expansion
  is highly anisotropic. The existence of these models shows that the
  observation of a highly isotropic cosmic microwave background
  temperature cannot alone be used to infer that the universe is
  close to a Friedmann-Lemaitre model.
\end{abstract}


\submitted

\maketitle

It is widely believed by cosmologists that the universe can be
accurately described by a Friedmann-Lemaitre (FL) model on sufficiently 
large scales. This belief stems from the fact that
the temperature of the cosmic microwave background (CMB) is measured
to be highly isotropic over the celestial sphere. In addition, certain
theoretical results lend credence to this 
notion. The fundamental theorem due to Ehlers, Geren \& Sachs (1968),
the so-called EGS theorem, states that if all fundamental observers 
measure an exactly isotropic CMB temperature during some time interval
in an expanding universe, then the universe is {\em exactly} an FL
model during this time interval. This theorem has been 
generalized by Stoeger \etal (1995), and subsequently by Maartens
\etal (1995a, 1995b), to the ``almost EGS theorem'', 
which states that if all fundamental observers measure the CMB
temperature to be almost isotropic during some time interval in an
expanding universe, then the 
universe is described by an almost FL model during this time
interval. In particular, the theorem implies that the overall 
expansion of the universe must be highly isotropic in the following
sense. The anisotropy in 
the expansion is described by the {\em shear scalar} $\Sigma$, defined
by
\begin{equation}
  \Sigma^2 = \frac{\sigma_{ab}\sigma^{ab}}{6H^2}\ ,
\end{equation}
where $\sigma_{ab}$ is the rate of shear tensor and $H$ is the Hubble
scalar (see Wainwright \& Ellis, p 18--19). The statement that the
expansion is highly isotropic means that $\Sigma\ll 1$.

For our purposes, the detailed form of the primary assumption of each
theorem is of importance. The EGS theorem requires the following
hypothesis:\\      

\begin{minipage}[l]{12cm}
  \Io {\em (the isotropy condition):}\\
  All fundamental observers measure the CMB temperature to be
  exactly isotropic during a time interval $I$.\\
  %defined by $\tE
  %\leq t \leq \tO$, where \tE is the time of last 
  %scattering and \tO is the time of observation.\\
\end{minipage}

\noindent
The almost EGS theorem requires the analogous hypothesis with
``exactly'' replaced by ``almost'':\\

\begin{minipage}[l]{12cm}
  \AIo {\em (the almost-isotropy condition):}\\
  All fundamental observers measure the CMB
  temperature to be almost isotropic during a time interval $I$.\\ 
\end{minipage}

\noindent
In the application of these theorems, the time interval $I$ is usually
identified with $\tEm \leq t \leq \tOm$, where \tE is the time of last 
scattering and \tO is the time of observation (see, for example,
Stoeger \etal (1995), p 1--2). Both theorems make use of kinetic
theory for describing the photons of the CMB (see, for example, Sachs
\& Ehlers (1971) for an introduction to relativistic kinetic
theory), and the proofs follow from the Einstein-Liouville
equations. The distribution function for the photon fluid is assumed
to be either exactly isotropic (the EGS theorem) or almost isotropic
(the almost EGS theorem). The multipoles of the CMB temperature
anisotropy can subsequently be written in terms of the anisotropies of
the distribution function (see equation (40) and (41) in Maartens
\etal (1995a)). It is also 
important to note that  
a number of technical assumptions about the behaviour of the
derivatives of the multipoles of the CMB temperature anisotropy must
be made in order to prove the almost EGS theorem (see, for example,
equation (1), (2), (3), and (4) in Maartens \etal (1995b)).

Although the EGS theorem and the almost EGS theorem
shed considerable light on the relation between idealized CMB
temperature observations and the geometry of cosmological models, they
cannot be used to conclude that the physical universe is close to an
FL model. The reason for this limitation is basically that the
hypotheses \Io and \AIo above include the stipulation ``during a
time interval $I$'', whereas we can only observe the CMB temperature
at 
one instant of time on a cosmological scale. This observational
limitation also implies that the technical assumptions needed for the 
almost EGS theorem are not observationally testable. In view of this
situation, 
it is of interest to ask what restrictions, if any, can be placed on
the anisotropy in the rate of expansion, using the following much weaker 
hypotheses:\\   

\begin{minipage}[l]{12cm}
  \It {\em (the weak isotropy condition):}\\
  All fundamental observers measure the CMB temperature to be
  isotropic at {\em some instant of time} \tO.\\ 
\end{minipage}

\begin{minipage}[l]{12cm}
  \AIt {\em (the weak almost-isotropy condition):}\\
  All fundamental observers measure the CMB temperature to be almost
  isotropic at {\em some instant of time \tO}.\\
\end{minipage}

\noindent
On the basis of continuity, we can argue that either of assumptions
\It and \AIt imply that all fundamental observers will
measure the CMB temperature to be almost isotropic in some time
interval $I_\delta$ of length $\delta$ centered on \tO. This 
time interval could, however, be much shorter than the time interval
$I$ referred to in \Io and \AIo. Nevertheless, if the
technical assumptions mentioned above concerning the almost EGS
theorem were satisfied, this theorem would imply that
during the time interval $I_\delta$ the model is close to an FL
model, and hence that the rate of expansion is highly isotropic ({\em
  i.e.} that $\Sigma \ll 1$). Thus, it is reasonable to ask whether
\It and \AIt  
imply that the overall expansion of the universe must be almost
isotropic at time \tO, without imposing {\em any} of the technical 
assumptions about the multipoles.  

In this Letter we answer this question in the negative. In particular
we show that {\em for a given time \tO,  there are spatially
  homogeneous 
cosmological models such that at \tO the CMB temperature is measured
to be isotropic by all fundamental observers, even though the overall 
expansion of the universe is highly anisotropic at \tO.}

We consider the class of non-tilted spatially homogeneous models whose
three-parameter isometry group is of Bianchi type VIII, and which are
also locally rotationally symmetric (LRS). Since we are modeling the
universe since the time of last scattering, it is reasonable to
consider models with dust as the source of the gravitational field. In
order to calculate the temperature of the CMB in these models, we 
adopt the approach of Nilsson \etal (1999)\footnote{From now on we
  will refer to this reference as NUW.}. This approach is based on 
the orthonormal frame formalism and makes use of expansion-normalized 
variables (see Wainwright \& Ellis (1997), p 112, for a motivation of
this normalization). We will use the combined gravitational
and geodesic evolution equations for models of Bianchi class A (which
includes Bianchi VIII), which are given by equations (24)--(30) in
NUW. This approach, which is purely geometric, differs from the
kinetic approach used in the EGS 
and almost EGS theorems in that effects of the CMB photons on the
gravitational field are neglected. These effects, however, will not
change the results in any significant way.

The temperature of the CMB in a non-tilted spatially homogeneous
cosmological model as function of angular position on the celestial
sphere is given by\footnote{This formula is only true for non-tilted
  spatially homogeneous models. It can, however, easily be generalized
  to tilted models. If the assumption of spatial homogeneity is
  dropped, the problem of finding the temperature of the CMB is
  significantly more complicated (see, for example, Dunsby (1997) for
  a covariant and gauge-independent approach).}
\begin{equation}
  T(\theta, \varphi) = T_{\rm e}\;{\rm exp}\left[ -
  \int_{\tauE}^{\tauO}(1+s)d\tau \right]\ , \label{eq:Tdef} 
\end{equation}
where
\begin{equation}
  s = \Sigma_{\alpha\beta}K^\alpha K^\beta \ , \label{eq:sdef}
\end{equation}
(see equation (58) in NUW). Here \TEt is the temperature
of the CMB at the surface of last scattering, which is assumed to be
the same for all photons of the CMB. The $K^\alpha$ are the direction
cosines of a particular null geodesic, and are related to the
usual spherical angles $\theta,\varphi$ according to 
\begin{equation}
  \label{eq:Kdefs}
  K^\alpha(\tauO) = \left( \sin\theta\cos\varphi\ , \
  \sin\theta\sin\varphi\ , \ \cos\theta \right) \ .
\end{equation}
The quantity
$\Sigma_{\alpha\beta}=\sigma_{ab}/H$ is the
expansion-normalized shear tensor, given by 
\begin{equation}
  \Sigma_{\alpha\beta} = {\rm diag}\;( \Sigma_+ +\sqrt{3}\Sigma_-\
 , \ \Sigma_+ -\sqrt{3}\Sigma_-\ , \ -2\Sigma_+ )\ .
  \label{eq:Sigmadef} 
\end{equation}
The time variable $\tau$ is the Hubble time, which is related to the 
cosmological clock time $t$ according to
\begin{equation}
  \frac{dt}{d\tau} = H^{-1}\ ,
\end{equation}
where $H$ is the Hubble variable. 

For LRS models, the shear is restricted by $\Sigma_-=0$, which, in
conjunction with (\ref{eq:Sigmadef}), implies that (\ref{eq:sdef})
simplifies to
\begin{equation}
  s = (1-3K_3^2)\Sigma_+ \ . \label{eq:LRSsdef}
\end{equation}
The geodesic equation for $K_3$ for LRS models simplifies to
\begin{equation}
  K_3' = 3(1-K_3^2)K_3\Sigma_+ \ , \label{eq:LRSgeoeq}
\end{equation}
(see equation (24h) in NUW). The prime denotes differentiation with
respect to $\tau$. Since neither (\ref{eq:LRSsdef}) nor 
(\ref{eq:LRSgeoeq}) involve the geodesic variables $K_1$ or $K_2$, we
need not consider the evolution equations for these variables. 
It follows from (\ref{eq:LRSgeoeq}) that if $K_3=0, \pm1$ initially,
it has these values throughout the entire evolution. We temporarily
exclude these values. 

To solve the differential equation (\ref{eq:LRSgeoeq}), we
introduce the new variable 
\begin{equation}
  \label{eq:ydef}
  y = \frac{1}{K_3^2} - 1\ ,
\end{equation}
which transforms (\ref{eq:LRSgeoeq}) into
\begin{equation}
  \label{eq:yprime}
  y' = -6\Sigma_+ y\ .
\end{equation}
It then follows that
\begin{equation}
  \label{eq:yquota}
                                %  \frac{y_{\rm e}}{y} = {\rm
                                %  e}^{6\beta} \ ,
  y(\tau)=y_{\rm e}\,{\rm e}^{-6\beta(\tau)}\ ,
\end{equation}
where
\begin{equation}
  \beta(\tau) = \int_{\tauE}^{\tau}\Sigma_+ d\tau \ ,
  \label{eq:betadef} 
\end{equation}
which we refer to as the {\em integrated shear}. Combining
(\ref{eq:ydef}) and (\ref{eq:yprime}) with (\ref{eq:LRSsdef}) leads to
\begin{equation}
  s = \frac{1}{6}\left( \frac{2}{y} - \frac{3}{1+y}
  \right)y' \ . 
\end{equation}
Integrating from \tauEt to \tauOt yields
\begin{equation}
  \int_{\tauE}^{\tauO}sd\tau = -\ln\left[ \left(\frac{y_{\rm o}
  +1}{y_{\rm e} +1} \right)^{1/2}\left(\frac{y_{\rm e}}{y_{\rm
  o}}\right)^{1/3}\right] \ .
\end{equation}
We note that $y_{\rm o}=\tan^2\theta$, which follows from
(\ref{eq:Kdefs}) and (\ref{eq:ydef}), and that $y_{\rm e}$ can be
expressed in terms of $y_{\rm o}$ using (\ref{eq:yquota}) with
$t=t_{\rm o}$. On substituting these results in (\ref{eq:Tdef}) we 
obtain 
\begin{equation}
  \label{eq:Tdef2}
  T(\theta,\varphi) = \TE\;{\rm e}^{-\Delta\tau} \frac{{\rm
      e}^{2\beta_{\rm o}}}{\sqrt{{\rm e}^{6\beta_{\rm o}} + (1-{\rm 
        e}^{6\beta_{\rm o}})\cos^2\theta}} \ ,
\end{equation}
where $\Delta\tau=\tauO-\tauE$ and $\beta_{\rm o} = \beta(\tauO)$. 
It is easily verified that (\ref{eq:Tdef2}) also holds for the
exceptional values $K_3=1,0,-1$, which correspond to
$\theta=0, \frac{\pi}{2},\pi$. Equation (\ref{eq:Tdef2}) shows 
explicitly that {\em in the present class of cosmological models, the
angular dependence of the CMB temperature is determined by the
integrated shear.} We emphasize that (\ref{eq:Tdef2}) holds
independently of the field equations. 

We now make use of the evolution equation for $\Sigma_+$ (equation
(24a) in NUW). With the LRS restrictions $\Sigma_-=0$ and $N_1=N_2$,
this equation simplifies to
\begin{equation}
  \label{eq:Sigmaeq}
  \Sigma_+' = -(2-q)\Sigma_+ - {\cal S}_+ \ ,
\end{equation}
where $q$ is the deceleration parameter, given by
\begin{equation}
  \label{eq:qdef}
  q = \frac12\left(1-\frac{1}{12}(N_3^2 - 4N_2N_3)\right) +
  \frac{3}{2}\Sigma_+^2 \ ,
\end{equation}
and
\begin{equation}
  \label{eq:spatialcurv}
  {\cal S}_+ = -\frac13(N_3^2 - N_2N_3)
\end{equation}
is a component of the anisotropic spatial curvature. For Bianchi VIII
LRS 
models, the spatial curvature variables satisfy $N_2N_3<0$, which
implies that ${\cal S}_+<0$. In addition, since $\Sigma_+^2\leq 1$, it
follows from (\ref{eq:qdef}) that $q\leq2$. As a result of these
inequalities, (\ref{eq:Sigmaeq}) implies that if $\Sigma_+\leq0$ then 
$\Sigma_+'>0$. Thus, any orbit that lies in the subset $\Sigma_+<0$ in
state space will eventually cross the surface $\Sigma_+=0$, and enter
the subset $\Sigma_+>0$. In other words, for an open set of
orbits\footnote{A detailed analysis of the orbits in fact shows that
  for almost all orbits, $\lim_{\tau\rightarrow-\infty} \Sigma_+ = -1$
  and $\lim_{\tau\rightarrow+\infty} \Sigma_+= \frac12$.} in the
state space, 
$\Sigma_+$ changes sign from negative to positive. It follows that for
any such orbit we can choose $\tau_0$, keeping $\Delta\tau$ constant,
so that the integrated
shear $\beta(\tau_{\rm o})$, as given by (\ref{eq:betadef}), is
zero. It follows from (\ref{eq:Tdef2}) that
\begin{equation}
  T(\theta, \varphi) = \TE\;{\rm e}^{-\Delta\tau}\ ,
\end{equation}
{\em i.e.}, that the temperature at this time is completely
isotropic. We 
summarize our results in the following theorem.
\begin{theo}
  There exist spatially homogeneous but anisotropic universes with the
  following properties:
  \begin{enumerate}
  \item the CMB temperature is completely isotropic at some instant of
    time \tO,
  \item at time \tO, the shear parameter $\Sigma_0$ satisfies
    $\Sigma_0\geq b$, where $b$ is a positive constant. 
  \end{enumerate}
\end{theo}

The theorem asserts that there are Bianchi VIII LRS models whose CMB
temperature evolve in the following way. As the universe expands after
last scattering, the CMB temperature reaches a state of maximum
anisotropy, then becomes completely isotropic at time $t_{\rm o}$ and
subsequently becomes increasingly anisotropic again. On the other hand
{\em the shear scalar $\Sigma$ becomes zero at some instant of time
  prior to $t_{\rm o}$}, and then increases again, exceeding some
value $b$ at time $t_{\rm o}$. A detailed
analysis of the state space shows that the value of the constant $b$
depends on the value of the density parameter $\Omega_{\rm o}$ at the
time of observation, and is not small unless $\Omega_{\rm o} \approx
1$, in which case the model is close to the flat FL model. The time
$t_{\rm o}$, which we regard as the present age of the universe, is
freely specifiable, since a point in state space determines the
physical time only up to a multiplicative constant (see Wainwright \&
Ellis, p 117). 

We emphasize that this
theorem does not contradict the almost EGS theorem, since the
technical assumptions on the multipoles are not satisfied, as we now
show. If $\tau$ is sufficiently close to $\tau_{\rm o}$ so that
$\beta(\tau)\ll 1$, the quadrupole component of the temperature
anisotropy can be written as $a_2 = \left| a_{20} \right|$, where 
\begin{equation}
  \label{eq:a2def}
  a_{20} = 4\sqrt{\frac{\pi}{5}}\left[ \beta - \frac{1}{7}\beta^2
  + {\cal O}(\beta^3) \right]\ .
\end{equation}
This result follows from (\ref{eq:Tdef2}) with $\beta_{\rm o}$
replaced by $\beta(\tau)$, using equation (62) in NUW.
By differentiating (\ref{eq:a2def}) with respect to $\tau$, and using
the result $\beta'(\tau)=\Sigma_+$, which follows from
(\ref{eq:betadef}), we obtain the following evolution equation for the
quadrupole moment $a_2$
\begin{equation}
  \label{eq:a2eq}
  a_{20}' = 4\sqrt{\frac{\pi}{5}}\left[1-\frac{2}{7}\beta + {\cal
  O}(\beta^2)\right]\Sigma_+\ . 
\end{equation}
The above equation is, to lowest order, equivalent to the evolution
equation for the quadrupole that arises in the kinetc theory based
approach of Maartens \etal (1995a) (differentiate equation (41) in
that paper with respect to $\tau$ and use equations (12) and (14) in
the same paper). From (\ref{eq:a2eq}) we see that when 
$\beta=0=\beta_{\rm o}$, {\em i.e.} when the temperature 
of the CMB is perfectly isotropic over the celestial sphere, the
derivative of $a_2$ is proportional to $\Sigma_+$. Therefore, since
$\Sigma_+$, in general, will not be small, the derivative of $a_2$
will not be small either. Hence the technical assumptions
of the almost EGS theorem are not satisfied.

It is also worth noting that the proof of the theorem does not depend
on the full set of field equations (equations (24a)--(24d) in NUW, with 
$\Sigma_-=0$ and $N_1=N_2$), but only on the fact that {\em the
  evolution equation (\ref{eq:Sigmaeq}) for the shear contains a
  driving term that 
has constant sign over the state space}, as does $-{\cal S}_+$. The
theorem thus applies to spatially homogeneous LRS cosmologies with
sources more complicated than dust. For example, in Bianchi I LRS
models with a homogeneous and source-free magnetic field, the evolution
equation for $\Sigma_+$ is of the form 
\begin{equation}
  \Sigma_+' = -(2-q)\Sigma_+ + 2{\cal H}^2\ ,
\end{equation}
where ${\cal H}$ is the expansion-normalized magnetic field.
                                %${\cal H}=\frac{h}{6H^2}$
Since 
this is of the same form as (\ref{eq:Sigmaeq}) and the geodesic
equation (\ref{eq:LRSgeoeq}) is exactly the same, the temperature of
the CMB for these models will be the same as in
(\ref{eq:Tdef2})\footnote{Since the Bianchi I LRS models with a
  homogeneous 
  magnetic field are closely related to the Bianchi II LRS models,
  which 
  are an invariant subset of the Bianchi VIII LRS models studied in
  this paper (see Collins (1972)), this result is not surprising.}.  
Thus there exist magnetic cosmologies of this type whose CMB
temperature is exactly isotropic at some time, but whose magnetic
field and shear is not small. In other words, an almost isotropic CMB
temperature cannot be used to limit a cosmological magnetic field,
unless additional assumptions about the nature of the multipoles are
made. This possibility seems to have been overlooked in Barrow (1997).

\section*{Acknowledgements}
The research was supported in part by a grant from the Natural
Sciences \& Engineering Research Council of Canada (JW),
G{\aa}l\"ostiftelsen (USN), Svenska Institutet (USN), Stiftelsen
Blanceflor (USN) and the University of Waterloo (USN, WCL). 

\References

\item[]
  Barrow J D 1997 Cosmological limits on slightly skew stresses {\em
    Phys. Rev.} D {\bf 55} 7451
  
\item[]
  Collins C B 1972 Qualitative magnetic cosmology {\em
    Commun. Math. Phys.} {\bf 27} 37

\item[]
  Dunsby P K S 1997 A fully covariant description of CMB anisotropies
  \CQG {\bf 14} 3391
  
\item[]
  Ehlers J, Geren P, and Sachs R K 1968 Isotropic solutions of the
  Einstein-Liouville equations \JMP {\bf 9} 1344

%%\item[]
%%  Ellis G F R, Matravers D R, and Treciokas R 1983 Anisotropic
%%  solutions of the Einstein-Boltzmann Equations: I. General formalism
%%  {\em Ann. Phys.} {\bf 150} 455
  
\item[]
  Maartens R, Ellis G F R, and Stoeger W R 1995a Limits on anisotropy
  and inhomogeneity from the cosmic background radiation {\em
    Phys. Rev.} D {\bf 51} 1525 
  
\item[]
  Maartens R, Ellis G F R, and Stoeger W R 1995b Improved limits on
  anisotropy and inhomogeneity from the cosmological background
  radiation {\em Phys. Rev.} D {\bf 51} 5942
  
\item[]
  Nilsson U S, Uggla C, and Wainwright J (NUW) 1999 A dynamical systems
  approach to geodesics in Bianchi cosmologies, to appear in {\em
    Gen. Rel. Grav.} (gr-qc/9908062)

\item[]
  Sachs R K and Ehlers J 1971 Kinetic theory and cosmology in ``{\em
    Astrophysics and General Relativity}'' 1968 Brandeis University
  Summer Institute in Theoretical Physics, Vol. 2, Chretien M, Deser
  S, and Goldstein J Eds. (Gordon \& Breach:London)
  
\item[]
  Stoeger W R, Maartens R, and Ellis G F R 1995 Proving
  almost-homogeneity of the universe: an almost Ehlers-Geren-Sachs
  theorem {\em Astrophys. Journ.} {\bf 441} 1

\item[]
  Wainwright J and Ellis G F R 1997 {\em Dynamical systems in
    cosmology} (Cambridge University Press: Cambridge)
\end{harvard}

\end{document}